\documentclass[]{spie}  

\usepackage{amsmath,amsfonts,amssymb}
\usepackage{graphicx}
\usepackage[colorlinks=true, allcolors=blue]{hyperref}
\usepackage{color}

\title{GRBAlpha: A 1U CubeSat mission for validating timing-based gamma-ray burst localization}

\author[a]{Andr\'as P\'al}
\author[b,a,c]{Masanori Ohno}
\author[a]{L\'aszl\'o M\'esz\'aros}
\author[d,c]{Norbert Werner} 
\author[b,d]{Jakub \v{R}\'{\i}pa}

\author[e]{Marcel Frajt}
\author[c]{Naoyoshi Hirade}
\author[e]{J\'an Hudec}
\author[e]{Jakub Kapu\v{s}}
\author[f]{Martin Koleda}
\author[f]{Robert Laszlo}
\author[g]{Pavol Lipovsk\'{y}}
\author[c]{Hiroto Matake}
\author[g]{Miroslav \v{S}melko}
\author[c]{Nagomi Uchida}

\author[a]{Bal\'azs Cs\'ak}
\author[h]{Teruaki Enoto}
\author[b]{Zsolt Frei}
\author[c]{Yasushi Fukazawa}
\author[i,b]{G\'abor Galg\'oczi}
\author[c]{Kengo Hirose}
\author[c]{Syohei Hisadomihi}
\author[k]{Yuto Ichinohe}
\author[a]{L\'aszl\'o L. Kiss}
\author[c]{Tsunefumi Mizuno}
\author[j]{Kazuhiro Nakazawa}
\author[l]{Hirokazu Odaka}
\author[c]{Hiromitsu Takahashi}
\author[c]{Kento Torigoe}

\affil[a]{Konkoly Observatory, Research Centre for Astronomy and Earth Sciences, Budapest, Hungary}
\affil[b]{Institute of Physics, E\"otv\"os Lor\'and University, Budapest, Hungary}
\affil[c]{Hiroshima University, Hiroshima, Japan}
\affil[d]{Department of Theoretical Physics and Astrophysics, Faculty of Science, Masaryk University, Brno, Czech Rep.}
\affil[e]{Spacemanic s.r.o., Bratislava, Slovakia}
\affil[f]{Needronix s.r.o., Bratislava, Slovakia}
\affil[g]{Faculty of Aeronautics, Technical University of Ko\v{s}ice, Slovakia}
\affil[h]{Hakubi Ctr. for Advanced Research, Kyoto Univ., Japan}
\affil[i]{Wigner Research Centre, Budapest, Hungary}
\affil[j]{Nagoya University, Nagoya, Japan}
\affil[k]{Rikkyo University, Tokyo, Japan}
\affil[l]{University of Tokyo, Tokyo, Japan}

\authorinfo{Further author information: (Send correspondence to A. P\'al)\\ A. P\'al: E-mail: apal@konkoly.hu, Telephone: +36-1-3919347}

\def\iic{I$^2$C}
 
\begin{document} 
\maketitle

\begin{abstract}
GRBAlpha is a 1U CubeSat mission with an expected launch date in the first half of 2021. It carries a $75\times75\times5\,{\rm mm}$ CsI(Tl) scintillator, read out by a dual-channel multi-pixel photon counter (MPPC) setup, to detect gamma-ray bursts (GRBs). The GRB detector is an in-orbit demonstration for the detector system on the Cubesats Applied for MEasuring and LOcalising Transients (CAMELOT) mission. While GRBAlpha provides 1/8th of the expected effective area of CAMELOT, the comparison of the observed light curves with other existing GRB monitoring satellites will allow us to validate the core idea of CAMELOT, i.e. the feasibility of timing-based localization.
\end{abstract}


\section{Introduction}
\label{sec:introduction} 

The detection and characterization of gamma-ray bursts is one of the most exciting fields in astronomy and it became even more prominent after the first detection of gravitational waves. However, simultaneous all-sky coverage and accurate localization are still challenging and this is where nano-satellites can play an essential role in the near future. 

One of such mission initiatives, Cubesats Applied for MEasuring and LOcalising Transients (CAMELOT\cite{werner2018}) aims to fill this gap by providing gamma-ray transient detection via a fleet of 3U CubeSats. The detector design for this mission is new in many aspects, including its full implementation at CubeSat level and the employment of SiPM detectors by Hamamatsu, called multi-pixel photon counter (MPPC). In this paper, we discuss the details of the in-orbit demonstration  mission called GRBAlpha, which will be launched in the first half of 2021. 

Compared to the final CAMELOT design, GRBAlpha will have a reduced effective scintillating surface area by a factor of $1/8$ on a single piece of caesium-iodine, while keeping two of the MPPC linear arrays as a dual-channel redundant readout detector -- where all of these, including the satellite platform electronics fit in an 1U CubeSat frame. In all other aspects, the payload side of GRBAlpha is identical to that of CAMELOT. Therefore, the comparison of the observed light curves provided by GRBAlpha with the existing gamma-ray observatories (Fermi, Swift) can validate the timing-based localization method\cite{ohno2018}, including estimates on the precision, accuracy and the spectral sensitivity. The higher level of integration and small size imply a simple attitude determination and control system (ADCS), however, we intend to test independent means of attitude recovery with a sub-degree accuracy.

The structure of this paper goes as follows. In Sec.~\ref{sec:detectordesign}, we detail our detector design which forms the payload of the satellite. The main features of the satellite platform is decribed in Sec.~\ref{sec:satellitedesign}. Our plans for mission operations are described in Sec.~\ref{sec:missionschedule} while we summarize our current status in Sec.~\ref{sec:summary}.

\begin{figure}[!t]
\begin{center}
\resizebox{15cm}{!}{\includegraphics{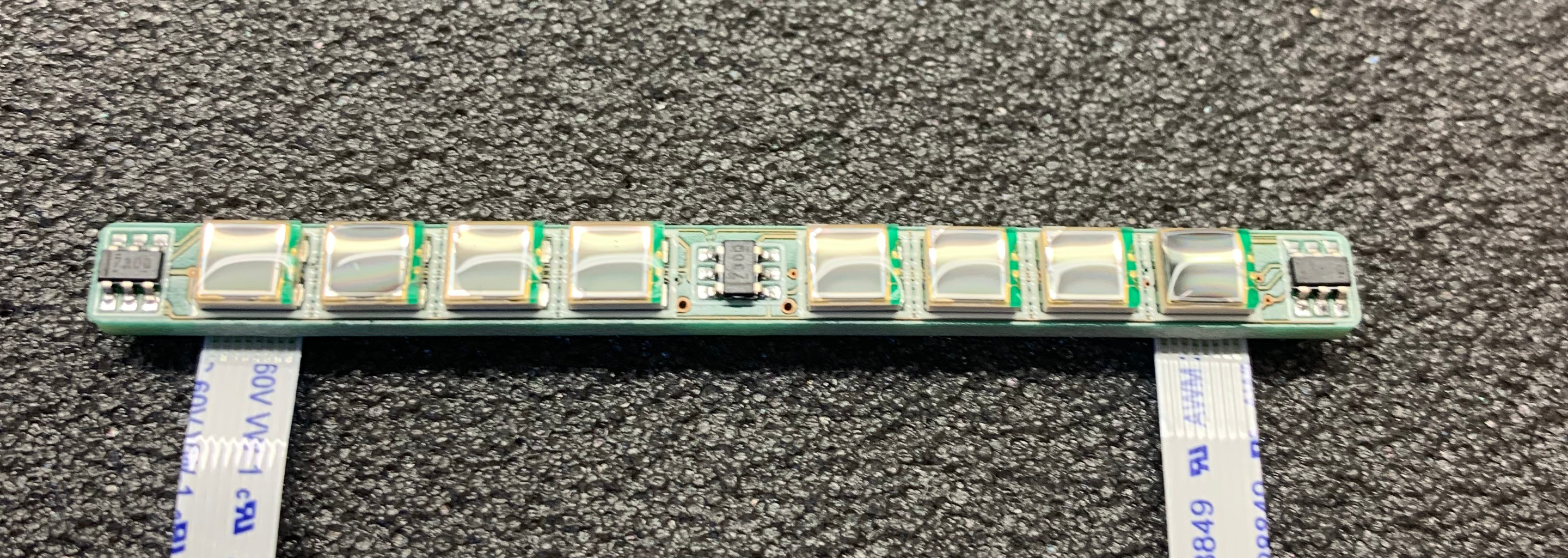}}\vspace*{2mm}
\resizebox{15cm}{!}{\includegraphics{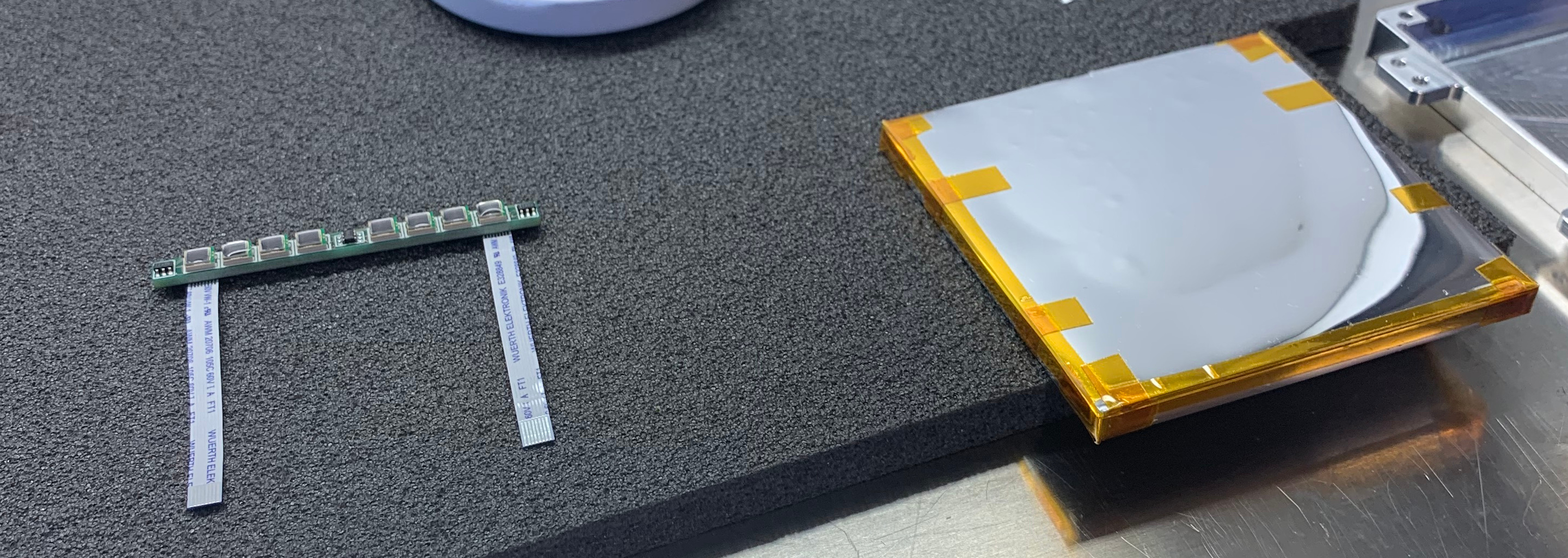}}\vspace*{2mm}
\resizebox{15cm}{!}{\includegraphics{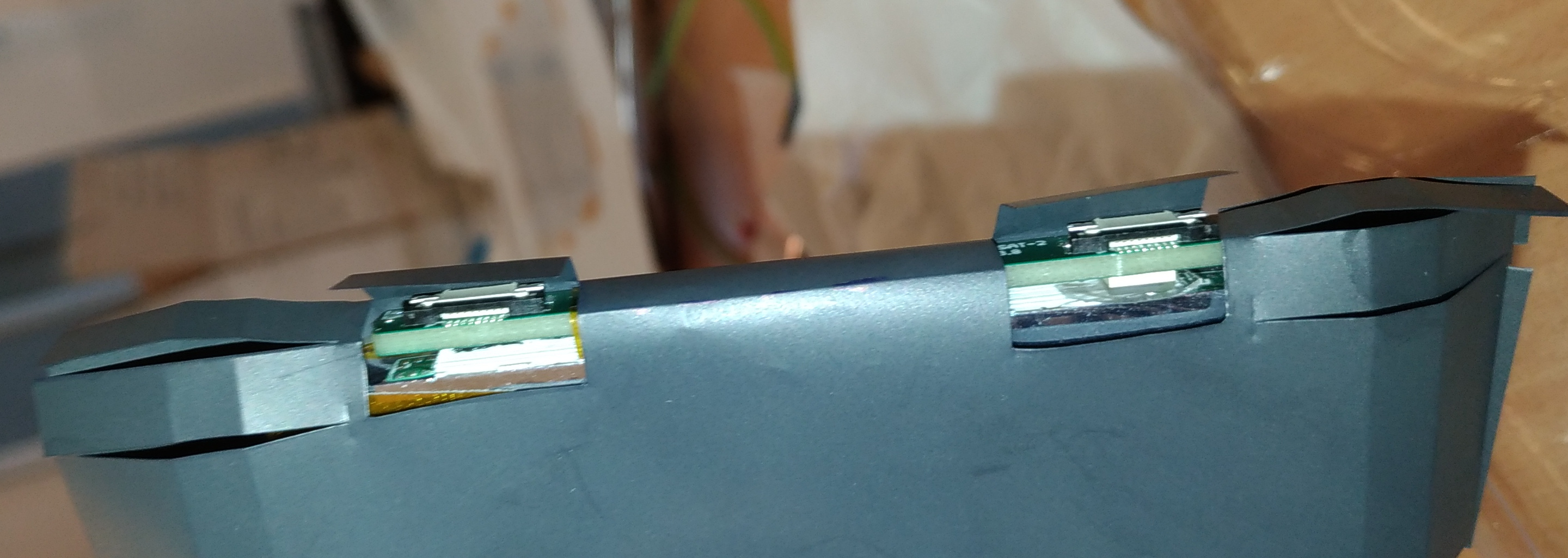}}\vspace*{4mm}
\caption{{\it Top:} The flight version of the $75\times5\,{\rm mm}$ PCB containing the $2\times 4$ MPPCs and the three thermometers. Optical glue has been applied to the MPPCs prior gluing it to the scintillator. {\it Middle:} The above MPPC board next to the scintillator, already wrapped into ESR foil, minutes before the permanent attachment. {\it Bottom:} Tedlar wrapping of the scintillator before attaching the flex flat cables.}
\label{fig:mppcboard}
\end{center}
\end{figure}

\begin{figure}[!t]
\begin{center}
\resizebox{15cm}{!}{\includegraphics{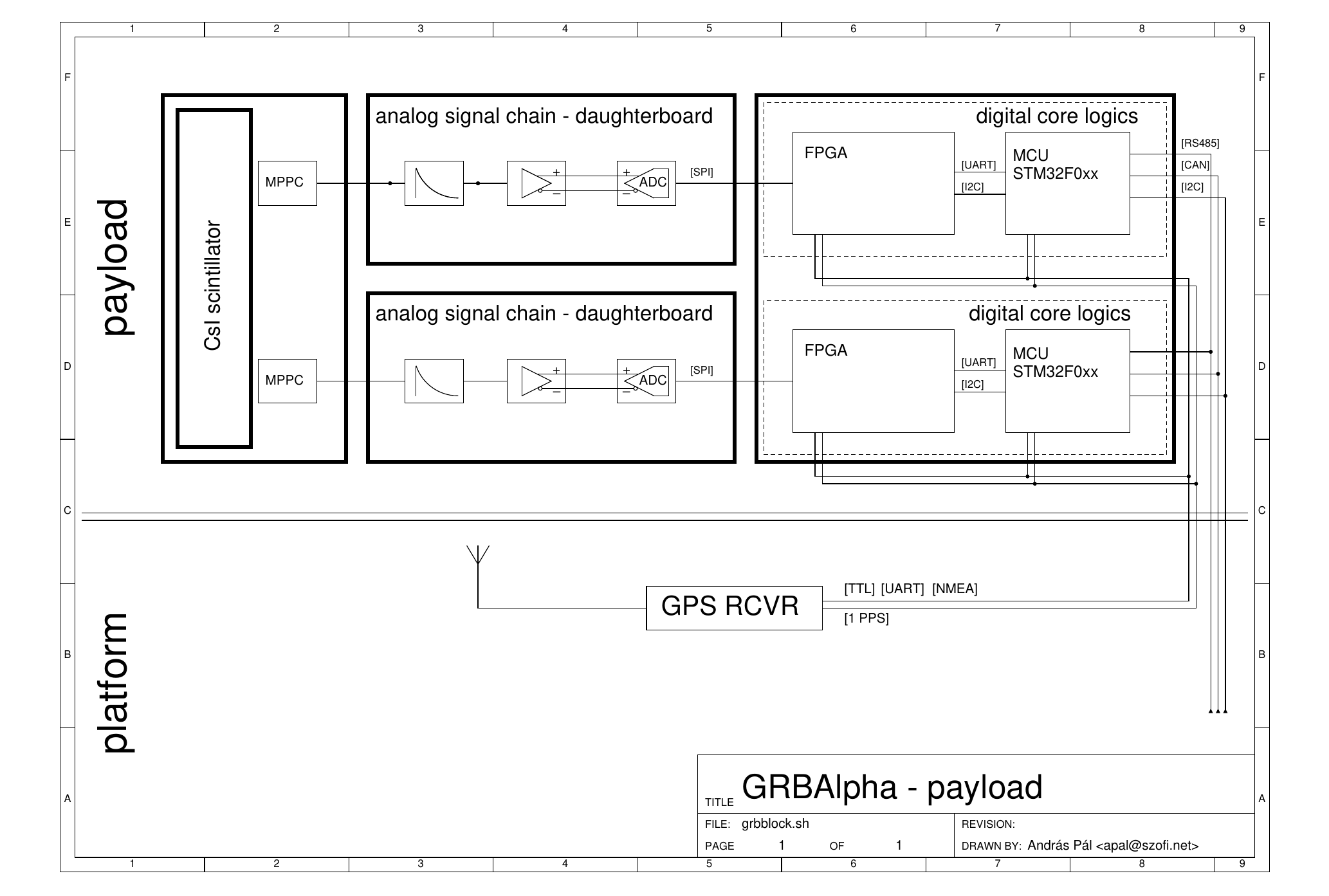}}\vspace*{4mm}
\caption{Block diagram of the GRBAlpha main scientific payload. In the  current realization, the payload contains two main components: the scintiallator case and the dual channel payload board. The payload board contains the so-called motherboard, supporting two identical digital blocks as well as two daughterboards mounted on the motherboard. To save space and weight, the connections between both the digital board and the daughterboards as well as between the daughterboards and the scintillator case are done by flat flex cables.}
\label{fig:grbblock}
\end{center}
\end{figure}

\begin{figure}[!t]
\begin{center}
\resizebox{15cm}{!}{\includegraphics{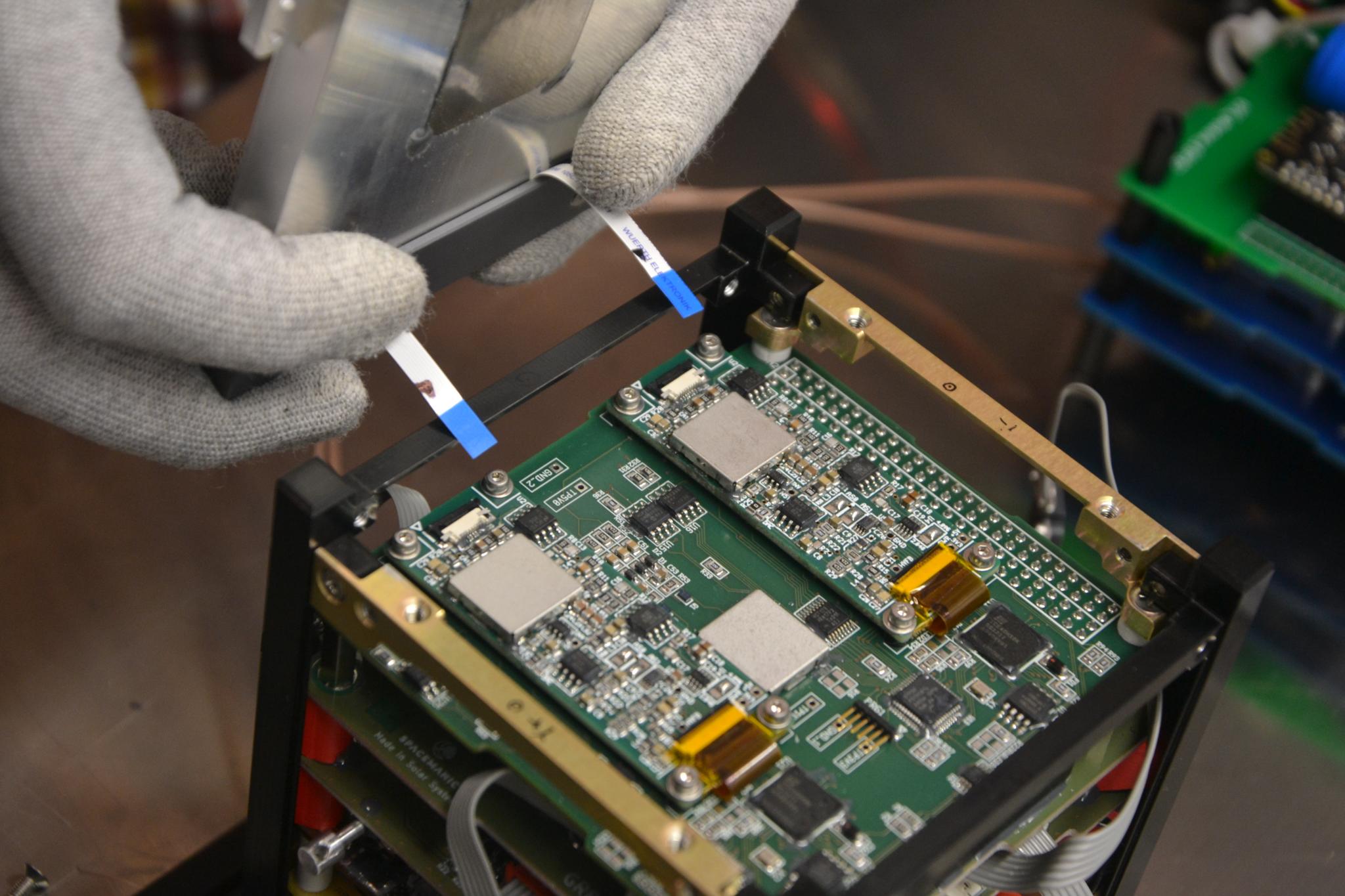}}\vspace*{4mm}
\caption{The components of the GRBAlpha main scientific payload just moments before the final assembly. The payload board is the topmost PCB on the  PC/104 CubeSat stack, supporting the two identical analog daughterboards into which the cables coming from the scintillator case are just being connected on this photo. The core logic (MCUs and FPGAs) are on the rightmost side of the payload PCB. }
\label{fig:grbpayloadassembly}
\end{center}
\end{figure}

\section{Detector design}
\label{sec:detectordesign}

The heart of the gamma-ray detector is a thallium activated cesium-iodine crystal with a size of $75\times75\times 5\,{\rm mm}$. This crystal is wrapped in a reflective foil - 3M Enhanced Specular Reflector (ESR). Attached to its side is a small PCB with a size of $75\times 5\,{\rm mm}$ supporting two linear arrays of 4 surface mounted multi-pixel photon counters. Therefore, these two arrays of MPPCs observe the same crystal, providing an increased sensitivity as well as redundancy. In addition, three small thermometers are attached to this PCB, providing a continuous monitoring of the interior of the detector via \iic{} interface. The whole setup is then packed into black tedlar (DuPont TCC15BL3) wrapping before fitting into the aluminium support case. The wrapping is designed to avoid any kind of light leak into the scintillator assembly (see Fig.~\ref{fig:mppcboard}).

The signal from the MPPCs runs through the analog signal chain formed by the pre-amplifier and the shaping amplifier before connected to the analog-digital converter (ADC) while the MPPC bias voltage is generated by an adjustable high voltage power supply (HV, for short). The amplifiers, ADCs, the HV supply and the digital-analog converter (DAC) responsible for adjusting the HV supply are located in a designated PCB atop the main payload electronics board and referred as daughterboards for simplicity. The aforementioned design implies that the interface between the daughterboards and the payload board is purely digital and also contains pass-through lines allowing a direct access of the MPPC board within the scintillator directly from the main payload board.  

The analog daughterboards are supported by the payload board which contains two identical digital cores, three hot-redundant communication interfaces with the satellite platform and the interfacing with the UART RX and PPS lines (provided by the on-board GPS receiver via the stacking connectors). In addition, the digital cores also support local storage arrays of $4+4$ identical 128Mb and 4Mb NOR flash and ferroelectric RAM (FRAM) chips, respectively, allowing a total space of $128+4$ megabytes for the full payload. The block diagram of the payload electronics can be seen on Fig.~\ref{fig:grbblock}. The components of the final detector assembly are displayed in Fig.~\ref{fig:grbpayloadassembly}. 

The digital core of the payload is formed by a 32-bit ARM Cortex-M0 from the STM32F0xx microcontroller unit (MCU) family and a field-programmable gate array (FPGA) from the iCE40HX family. The ADC on the analog daughterboard is connected directly to the FPGA, providing fast and continuous readout of the channels along with good flexibility on the digital signal processing. This advanced ADC control is interfaced with an 8-bit AVR architecture CPU\footnote{https://en.wikipedia.org/wiki/AVR\_microcontrollers} which is implemented within the FPGA as a soft processor core\footnote{https://opencores.org/projects/softavrcore}. The ADC control core is interconnected to the CPU via the 6-bit wide address bus and 8-bit wide data bus of the AVR I/O peripheral interface. By default, the ADC employs 8 registers (out of the total 64) of the peripheral register map, however, it is also possible to map the counter array of the ADC peripheral directly to the CPU virtual memory when the ADC peripheral core runs in histogram mode. 

The operation of the FPGA is supervised by the ARM microcontroller via UART and an optional \iic{} interface. The microcontroller can also actively re-program the FPGA bitstream FRAM if needed. The MCU is also responsible for the implementation of the hot-redundant communication interfaces with other components of the satellite, using RS485, \iic{} and CAN interfaces, at the same time. We note, that hot-redundancy means here the possibility to simultaneously {\it accept} packets from all of the communication lines, but the outgoing interface is the one preferred by the packet router. On higher layers, the communication interface is based on the CubeSat Space Protocol (CSP\footnote{https://en.wikipedia.org/wiki/Cubesat\_Space\_Protocol}), where switching between the various physical interfaces is performed by the routing capabilities of CSP. With the exception of the CSP address itself, the firmware running on the two MCUs and the bitstream loaded in the configuration FRAM of the FPGAs are strictly identical. Moreover, the flash area of the MCU is split into three parts: a bootloader, a configuration area and a main (scientific) firmware. The bootloader is capable to re-program and verify the main firmware as well as to perform a basic level health check of the payload electronics. Scientific data acquisition is managed by the main firmware. The aforementioned design also implies that the FPGA is not connected to the satellite bus, however, it is possible to do so using the CSP routing features of the MCU. 

\begin{figure}[!t]
\begin{center}
\resizebox{15cm}{!}{\includegraphics{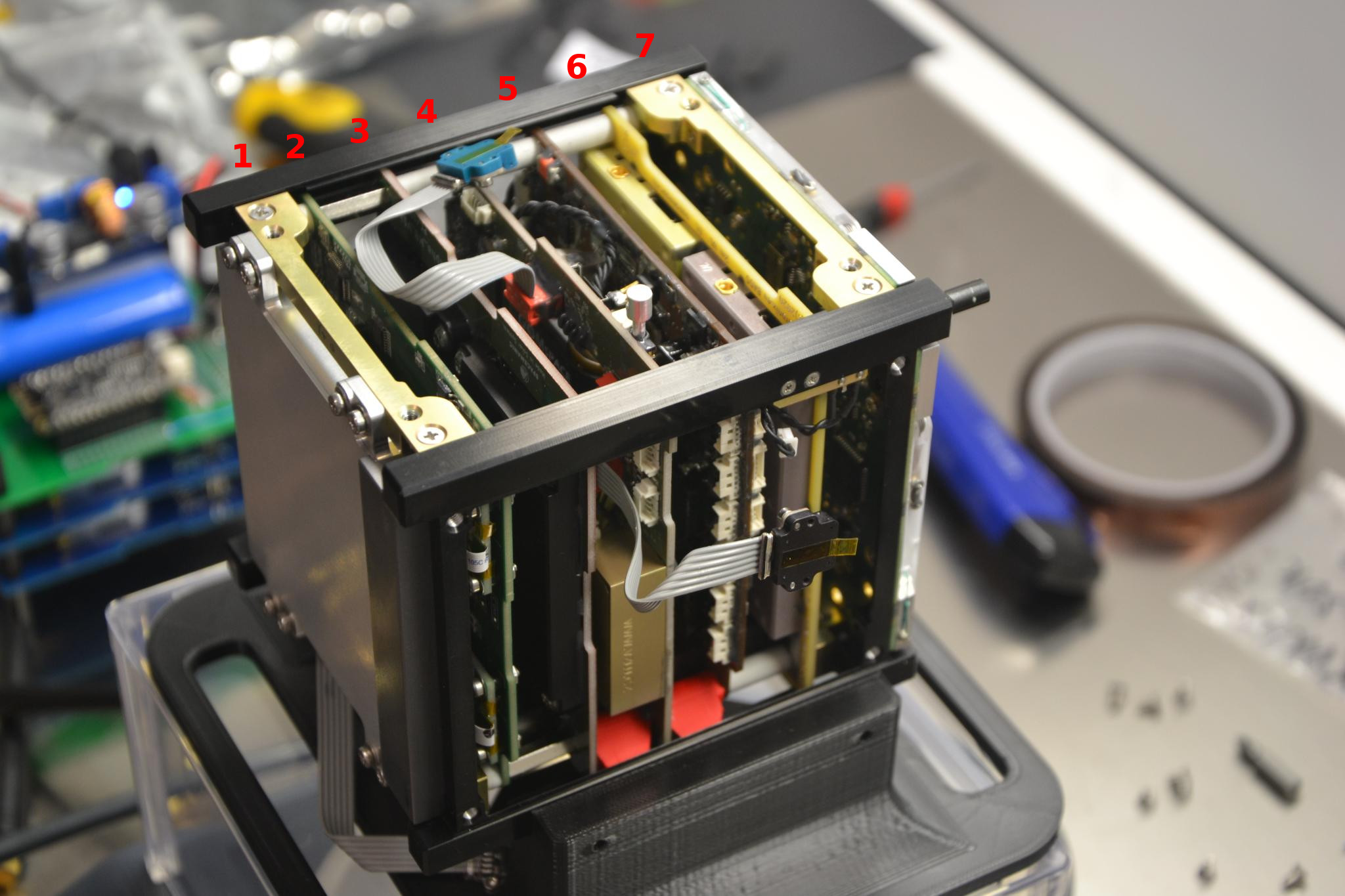}}\vspace*{4mm}
\caption{The internal components of the GRBAlpha 1U satellite. From $Z+$ to $Z-$, i.e. from left to right on this image: 1. the scintillator case; 2. the dual channel payload board; 3. the board supporting the on-board computer and GPS receiver; 4. sensor board. 5. power supply unit, 6. dual UHF + VHF transceiver module 7. deployable antennas. }
\label{fig:grbalphainterior}
\end{center}
\end{figure}

\section{Satellite design}
\label{sec:satellitedesign}

\begin{figure}[!ht]
\begin{center}
\resizebox{8cm}{!}{\includegraphics{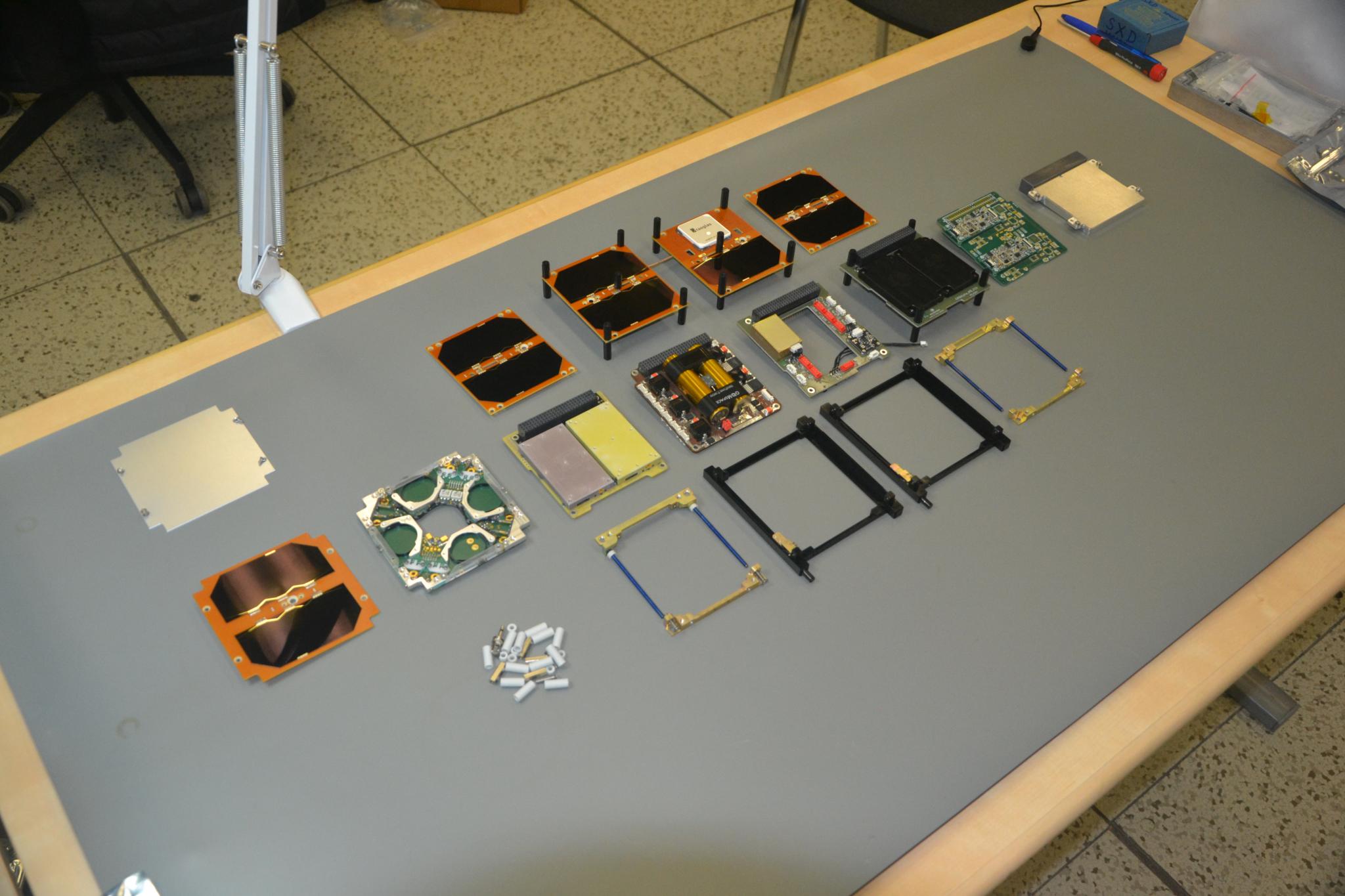}}\hspace*{4mm}%
\resizebox{8cm}{!}{\includegraphics{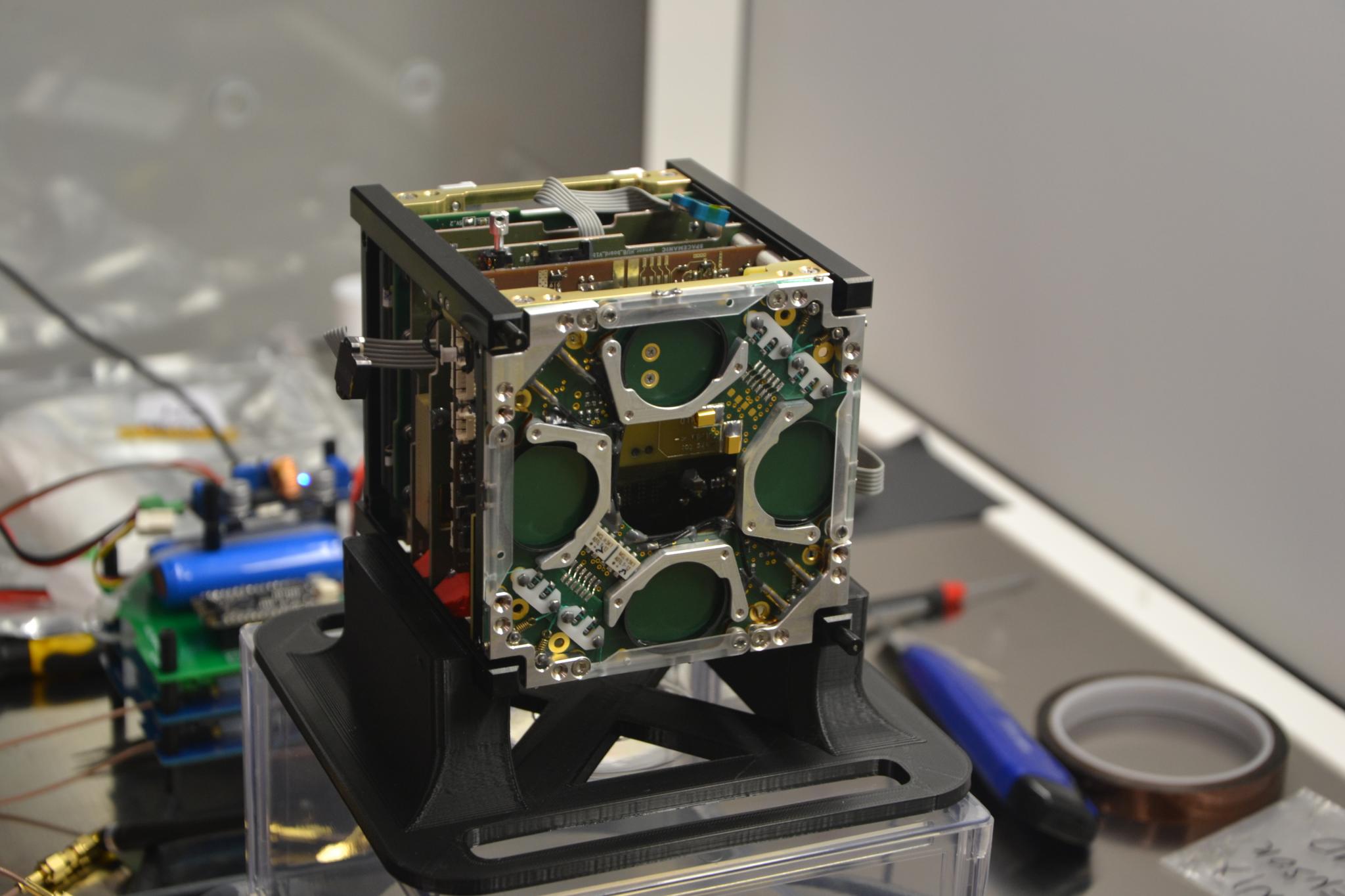}}\vspace*{4mm}
\resizebox{8cm}{!}{\includegraphics{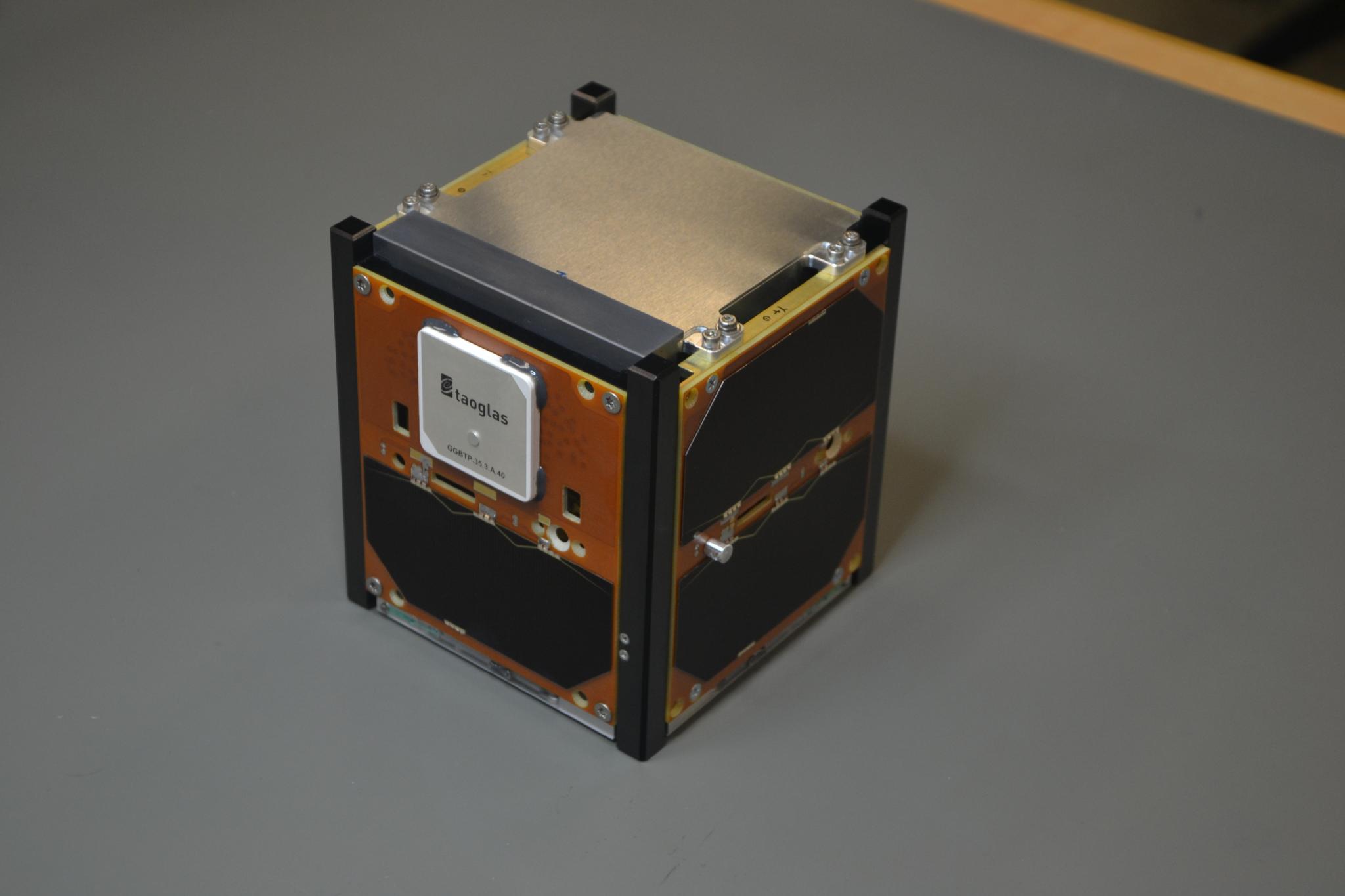}}\hspace*{4mm}%
\resizebox{8cm}{!}{\includegraphics{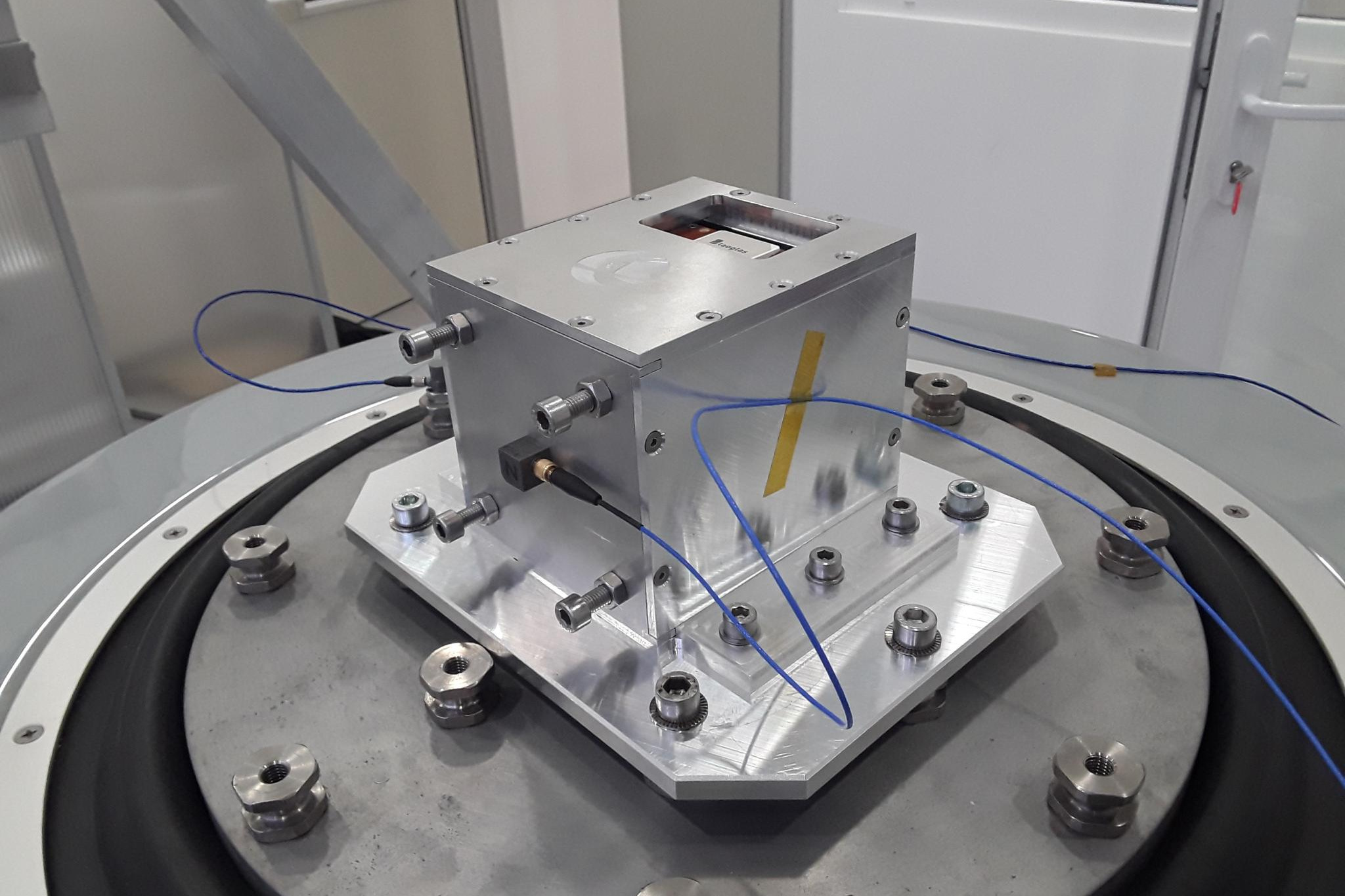}}\vspace*{4mm}
\caption{{\it Top left:} The components of the GRBAlpha 1U satellite before starting the integration. {\it Top right:} Bottom ($Z-$) view of the satellite (without the bottom solar panel), showing the antenna deployer. {\it Bottom left:} the full assembly of GRBAlpha with detector on the top side and GPS antenna on the left side. {\it Bottom right:} GRBAlpha before starting the vibration tests. }
\label{fig:grbalphacomponents}
\end{center}
\end{figure}

As we implied earlier in Sec.~\ref{sec:detectordesign}, the satellite employs the CubeSat Space Protocol for performing internal communications between the various payload and platform components. The platform-side components include the UHF and VHF transceivers (by Needronix), the power supply unit (by GOMspace), the on-board computer and GPS receiver (by Spacemanic; for GPS time-stamping see P\'{a}l et al. 2018\cite{pal2018}), and a sensor board with the sun-sensors (by Needronix), the space X-ray dosimeter (SXD by VZLU), magnetometers, gyroscopes, and thermometers. These internal components are displayed in Fig.~\ref{fig:grbalphainterior}. The satellite features permanent magnets for the passive detumbling of the satellite within a few orbits. However, attitude information is required in order to properly interpret the scintillator data. For this purpose, small sun sensors are attached in the four sides ($X+$, $Y+$, $X-$ and $Y-$) of the satellite and an inertial measurement unit (IMU) is also available on-board. The components of the satellite are shown in Fig.~\ref{fig:grbalphacomponents} at the various stages of the integration as well as prior vibration tests. 

The individual boards of the stack of the flight electronics are connected via the CubeSat standard PC/104 connector. This connector provide the distributed power supply, the internal communication interfaces, the GPS receiver lines, global master reset and various auxiliary functionality at the same time. Redundancy is applied in the internal communication using the aforementioned CSP library and the physical interfaces of RS485, two \iic{} and CAN buses. However, not all of the components support all of these physical interfaces at the same time. In order to further increase the reliability of the internal communication packet routing is applied between interfaces which are not connected with the same physical lines. Routing rules can either be applied and/or changed temporarily and permanently, depending on the health of the components in the telemetry transceivers, in the on-board computer, in the GPS module as well as in the two independent units of the payload. The attitude sensors are attached to the platform components only by a single \iic{} channel, however, the redundancy here is implied by the presence of numerous, identical detectors. 

\section{Mission schedule and expected operations}
\label{sec:missionschedule}

The launch of GRBAlpha is expected in the first half of 2021 on a Soyuz rocket from Baikonur. The expected mission lifetime is at least one year. After the launch the satellite will undergo detumbling - when a satellite is separated from the deployer, it tumbles freely and a time period of a few orbits will be need for the satellite to get stabilized via permanent neodymium magnets in special 3D printed holders. After the satellite is stabilised, the magnets will point their poles to the opposite magnetic poles of the Earth. This will cause the satellite to oscillate during each orbit. To reduce the amplitude of such oscillation, plates of soft magnetic amorphous material are glued to the back sides of the solar panels. 
Furthermore, during the fist passes after the launch the health of the main payload and other satellite sub-systems will be checked through housekeeping data e.g. temperatures, voltages and through the detected gamma-ray spectrum and the received GPS signal.

After these initial post-launch operations the satellite will start its normal operations. Besides being in the bootloader, normal in-orbit operations consist of three modes: the science data taking mode (SCI mode), the calibration mode (CAL mode), and the standby mode (STBY mode). The main operation mode is the SCI mode, where the given energy-binned histogram data is accumulated and it is sent to the MCU and OBC at a certain frequency, which corresponds to the time binning of the histogram data. During the continuous data taking phase, these energy and time binning factor should be coarse to save the data size, but once the on-board trigger algorithm detect the sudden increase of the observed count rate caused by astronomical transient such as GRBs, this binning factor will be switched to a finer value to start the trigger data taking phase, where detailed light curves and spectra for a given data size and/or duration are accumulated. Then, the binning parameter returns to the value for the continuous data taking phase. 

\begin{figure}[!ht]
\begin{center}
\resizebox{15cm}{!}{\includegraphics{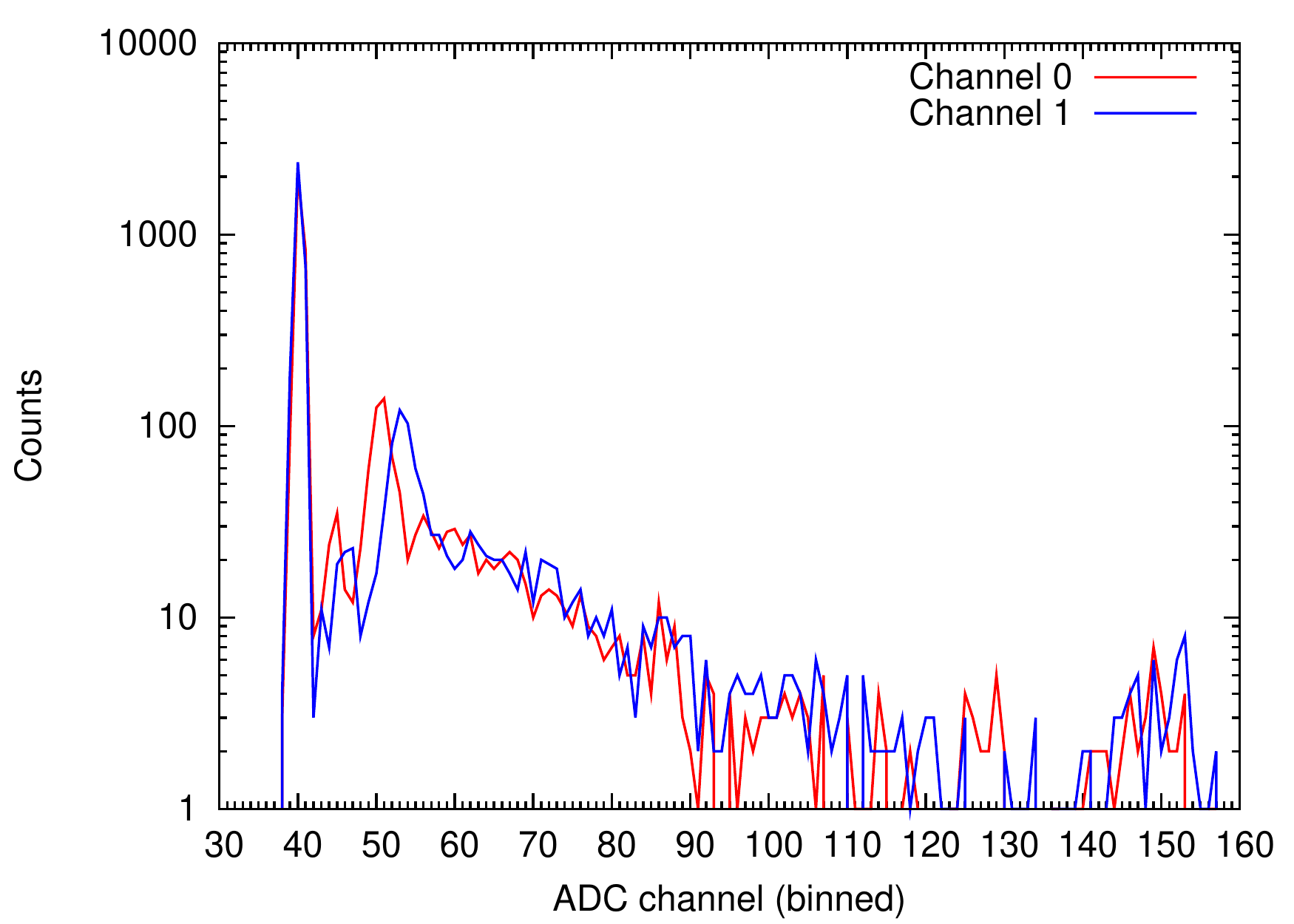}}\vspace*{4mm}
\caption{Calibration spectra of the scintillator detector taken after the vibration tests. The characteristics peaks of the 26.4\,keV and 59.54\,keV Am-241 lines are clearly seen at the ADC bins of $\sim 45$ and $\sim 53$, just slightly above the pedestial at $\sim 40$. As of this writing, the calibration is still underway with other gamma-ray sources, such as Ba-133.}
\label{fig:calib}
\end{center}
\end{figure}

The binning parameters can be controlled by commands to start the CAL model. In this mode, various types of calibration data are expected to be accumulated (e.g., finer energy binning spectrum to monitor the characteristic lines in the background to check the detector gain), in a similar fashion what is performed during the ground calibration of the energy channels (see e.g. Fig.~\ref{fig:calib} for an example). After such calibration measurements are finished, the detector is switched back to the continuous data taking mode by setting nominal binning parameters.

\begin{figure}[!t]
\begin{center}
\resizebox{15cm}{!}{\includegraphics{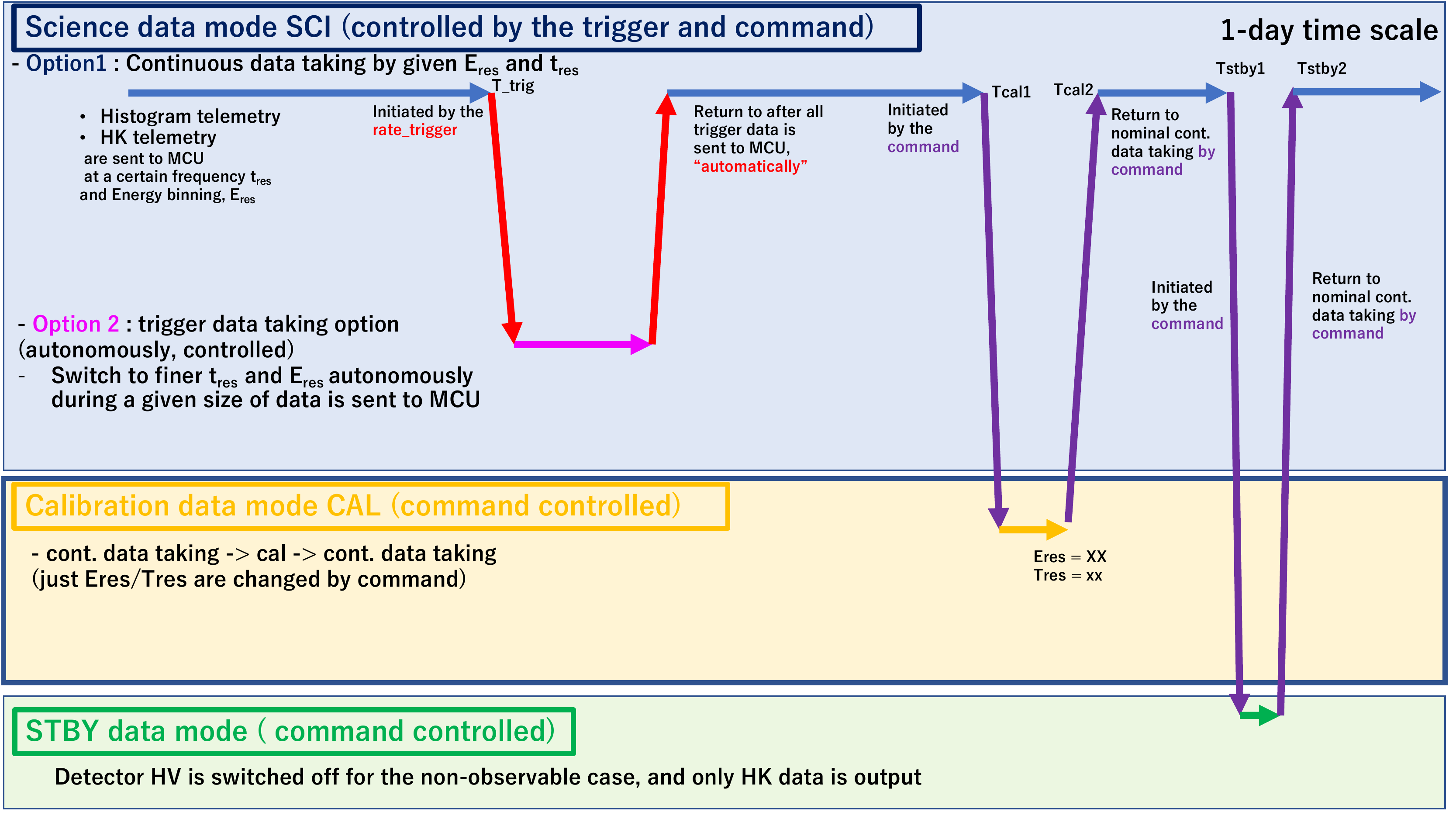}}\vspace*{4mm}
\caption{The flow chart of the expected operation plan for the GRBAlpha. Three modes; the science data taking mode, the calibration mode and the standby mode are switched by the command.}
\label{fig:datamode_flow}
\end{center}
\end{figure}

The STBY mode is for unpredictable situations, when the high-voltage needs to get switched off, but the housekeeping data are read out to monitor the status of the payload. Figure \ref{fig:datamode_flow} shows the flow chart of the expected operation plan.

\section{Summary}
\label{sec:summary}

GRBAlpha will be one of the first CubeSats missions to perform the monitoring of the high-energy sky. It will be an in-orbit demonstration for the detector concept of the larger CAMELOT constellation\cite{werner2018}, which is expected to consist of at least 9 satellites. A number of other CubeSat missions is under development, including the HERMES\cite{fiore2020,evangelista2020} constellation, BurstCube\cite{racusin2017}, GRID\cite{wenjiaxing2019}, etc., indicating that in the next decade constellations of CubeSats providing both all-sky coverage and localisation capability will be highly complementary to large missions monitoring the high energy sky.

\acknowledgments 
This project has primarily funded by the grant KEP-7/2018 of the Hungarian Academy of Sciences. Additional support is received by the European Union, co-financed by the European Social Fund (Research and development activities at the E\"{o}tv\"{o}s Lor\'{a}nd University's Campus in Szombathely, EFOP-3.6.1-16-2016-00023). The work of A.P. and L.M. was supported by the GINOP-2.3.2-15-2016-00033  project which is funded by the Hungarian National Research, Development and Innovation Fund together with the  European Union. We are grateful to RMC s.r.o. for providing the solar cells for the mission, and for the help with the design of the solar panels and the related measurements and simulations. 

\bibliography{report} 
\bibliographystyle{spiebib} 

\end{document}